\documentstyle[preprint,aps,prb,psfig]{revtex}
\begin{document}
\def\alt{\mathrel{\mathpalette\vereq<}}
\def\agt{\mathrel{\mathpalette\vereq>}}
\def\ep{{\epsilon}}
\def\tg{{\tilde{\cal G}}}
\def\bk{{{\bf k}}}

\title{Infrared absorption in superconducting MgB$_{2}$}

\author{C.S.Sundar, A. Bharathi, M. Premila, T.N. Sairam, 
S. Kalavathi, G.L.N. Reddy,
V.S. Sastry,  Y.Hariharan and T.S. Radhakrishnan}

\address
{
Materials Science Division,
Indira Gandhi Centre for Atomic Research\\
Kalpakkam, India 603 102
}

\maketitle

\begin{abstract}
{
Infrared absorption measurements in the range of 125 to 700 cm$^{-1}$
have been carried out as a function of temperature upto 5 K
in MgB$_{2}$. The absorption spectrum is
characterised by a broad band centred at 485  cm$^{-1}$, with
shoulders at 333 and 387 cm$^{-1}$. Studies on the temperature
dependence of absorption,  indicate
that these modes initially harden with the lowering of
temperature, and this trend is arrested at $\sim $ 100 K, below
which they soften.  Further, in the case of the mode at 333 cm
$^{-1}$, there is a distinct softening associated with the
superconducting transition at 39 K.  The implications of these
experimental results in the context of superconductivity in
MgB$_{2}$ are discussed.  } \\

\noindent {PACS numbers: 74.70.Ad, 74.25.Kc, 78.30.-j \\}
\end{abstract}

\section{Introduction}

The recent discovery of superconductivity \cite{akimitsu} at 39 K
in the binary
intermetallic MgB$_{2}$, having a simple
hexagonal structure consisting of alternating honeycomb layers
of B and closed packed layers of Mg, has
evoked a widespread interest.  Initial experiments suggest that the
superconductivity in this system arises due to phonon-mediated
interaction, as supported  by experiments on isotope effect
\cite{budko}; measurements on the superconducting gap by
tunnelling \cite{schmidt} and optical spectroscopy
\cite{gorshunov}, as also the decrease of T$_{c}$ with the application
of pressure \cite{chu}. There have been several theoretical
calculations \cite{kortus,pickett,andersen,satta} that emphasise
the importance of electron-phonon coupling, though alternative
mechanisms \cite{hirsch,baskaran} for superconductivity have also
been proposed. Electronic structure 
calculations in  MgB$_{2}$ indicate that Mg is completely ionised and
the bands at the Fermi level are derived from $\sigma $  orbitals
of boron. There are four distinct zone centre vibrational modes:
a silent mode B$_{1g}$,  the  doubly degenerate Raman mode,
E$_{2g}$, and two infrared active modes of  A$_{2u}$ and
E$_{1u}$ symmetry.  While there is a general agreement with
regard to electronic structure and vibrations, the details
differ, both with respect to the calculated frequencies and the
relative importance of these modes with respect to
superconductivity in the system. For example, in the
calculations of An and Pickett \cite{pickett}, using the
deformation potential approach, it is the E$_{2g}$ mode that is
shown to have a dominant coupling with the electrons, whereas in
the calculations of Kong et al \cite{andersen}, the electron
phonon interaction is spread over all the modes.  Inelastic
neutron scattering experiments measuring the vibrational density
of states \cite{osborne,sato,yildrim} indicate acoustic modes at
less than 40 meV and peaks in the phonon density of states at 
54, 78, 89 and 97 meV 
corresponding to the optic modes.  Raman scattering measurements
\cite{bohnen,chen,goncharov} indicate a mode at $\sim $ 600
cm$^{-1}$, which is characterised by very large width $\sim $
200 cm$^{-1}$, indicative of strong electron phonon interaction.

In this paper, results of infrared  absorption measurements on
superconducting MgB$_{2}$, covering a spectral range of 125 -
700 cm$^{-1}$, in which the optic modes are predicted to exist,
are reported. In addition supportive experiments have also been
carried out in the mid infrared range, upto 2000 cm$^{-1}$.
Based on these experiments and comparison with theoretical
calculations \cite{kortus,pickett,andersen,satta}, the optic
modes of MgB$_{2}$ are identified. From studies on the
temperature dependence of infrared absorption, it is seen that
the optic modes show an interesting temperature variation in
that the hardening behaviour at low temperatures is arrested
below 100 K, below which they soften. In addition, the IR mode
at 333 cm$^{-1}$ shows a distinct softening below T$_{c}$.

\section{Experimental Details}

The MgB$_{2}$ sample used in the present experiments was
prepared from  Mg (99.99 $\% $) powder of 50 mesh and B (99.98
$\% $) powder of 325 mesh. The starting materials were
thoroughly mixed and put in Ta tube that was sealed in Ar
atmosphere. This was subsequently sealed in a quartz tube that
was heat treated  at 1223 K for 2 hours. After cooling down to
room temperature, the polycrystalline lumps were crushed using
agate mortar and pestle and subsequently used for
experimentation. X - ray diffraction measurements were carried
out using Cu-K$_{\alpha } $ radiation in the Bragg-Brentano
geometry. AC susceptibilty measurements were carried out with a
mutual inductance bridge and lock-in amplifier, using a dipstick cryostat operating
in the 4 to 300 K range. The diamagnetic signal corresponding to
the sharp superconducting transition at 39 K, and the results of the
x-ray diffraction measurements are shown in Fig.1.  The
diffraction pattern of  MgB$_{2}$ can be indexed to hexagnal
structure (P6/mmm) with lattice parameters: a= 3.0864 A and
c=3.5253 A, in agreement with earlier studies \cite{jorgensen}.
The diffraction pattern of crystalline $\beta $ rhombohedral 
boron is also
shown, and it is seen that boron peaks are not discernable in the
diffraction pattern of MgB$_{2}$.  However, there are a few
peaks that could be associated with MgO, whose concentration is
estimated to be $\sim 1 \% $. From the powder ac-susceptibilty
measurements, using Pb powder as standard, the superconducting
volume fraction was estimated to be $75 \% $.

Infrared absorption measurements were carried out  on finely
ground MgB$_{2}$ sample pelletized along with KBr,  using a
BOMEM -DA8 spectrometer operating with a resolution of 2
cm$^{-1}$.  Measurements in the range of 125 - 700 cm $^{-1}$
were carried out using a mylar beam splitter and a DTGS
detector. Experiments in the mid infrared range of 400 to 2000
cm$^{-1}$ have been carried out using the combination of KBr
beam splitter and MCT detector. To study the temperature
variation of IR modes, the sample was mounted inside a JANIS
continuous flow cryostat in which temperature variation of  300
to 5 K could be achieved.

\section{Results and Discussion}

Fig.2 shows the IR absorbance of MgB$_{2}$ in the range of 300
to 650 cm$^{-1}$ . The region below 300 cm$^{-1}$ is supressed,
as it is dominated by KBr absorption. The absorption spectrum in
MgB$_{2}$ is characterised by a broad band centred at 485
cm$^{-1}$ with shoulders at  333, 387 cm$^{-1}$.  Further sharp
features are also noted at  542, 592 and 633 cm$^{-1}$.  The
latter modes match with those of $\beta $ rhombohedral boron 
\cite{nogi}, whose absorption spectrum is also shown. The occurrence of these 
modes, while the x-ray diffraction pattern does not indicate the presence 
of any B (cf. Fig. 1), p points to the activation of B-like modes in MgB$_2$
due to disorder (see below). Information on the
frequencies, widths and intensity of the various phonon modes
have been obtained by fitting  the absorption curves to a sum of
Gaussians and a linear background.  The resulting fits along
with the components is shown in the right panel of Fig.2, for
the two representative temperatures of 297 K and 5 K.

Factor group analysis predicts for MgB$_2$ ( space group P6/mmm,
z=1) B$_{1g}$ + E$_{2g}$ + E$_{1u}$+ A$_{2u}$  zone centre optic
modes, of which E$_{1u}$ and A$_{2u}$ are IR active and E$_{2g}$
is Raman active. There have been several calculations of these
mode frequencies \cite{kortus,pickett,andersen,satta,yildrim} with a
general agreement. Kortus et al \cite {kortus} have calculated
the two IR active modes of E$_{1u}$ and A$_{2u}$ symmetry to be
at 320 and 390 cm$^{-1}$ respectively and a Raman mode of
E$_{2g}$ symmetry at 470 cm$^{-1}$. In the calculations of Kong
et al \cite{andersen}, carried out using LMTO method, the
infrared modes are at 335 and 401 cm$^{-1}$ and the Raman mode
is at 585 cm$^{-1}$.  Yildrim et al \cite{yildrim} have
calculated the infrared modes to be $\omega$(E$_{1u}$) = 40.7
meV,  $\omega$(A$_{2u}$) = 49.8 meV and the Raman mode
$\omega$(E$_{2g}$) = 74.5 meV. Comparing these theoretical
calculations \cite{kortus,andersen,satta,yildrim} with our
experimental results, we identify the absorption features at 333
and 387 cm$^{-1}$ with E$_{1u}$ and A$_{2u}$ infrared modes.

As for the absorption band centred at 485 cm$^{-1}$, we first
note that it is different from the Raman mode identified to be
at 560 cm$^{-1}$ in the experiments by Bohnen et al
\cite{bohnen} and at 620 cm$^{-1}$ in the experiments of Chen et
al  \cite{chen}, and Goncharov et al \cite{goncharov}. In all
these experiments, the Raman mode is observed to be very broad
$\sim 200 $ cm$^{-1}$. Chen et al \cite{chen} have suggested that
this broad feature arises due to disorder  which relaxes the momentum
selection rule resulting in  phonons in the entire Brillouin
zone being sampled in the Raman experiment. Taking cue from
this, we note that the broad feature at 485 cm$^{-1}$, observed
in the present infrared absorption experiments (cf. Fig.2) can
be associated with the peak in the phonon density of states at
$\sim $ 54 meV, that is seen in the theoretical calculations and
neutron scattering experiments\cite{osborne,yildrim}. In effect
the broad absorption band centred at 485 cm$^{-1}$ arises
due to sampling of the phonons in this energy range over the
entire Brillouin zone. The exact nature of disorder that is
being invoked to account for  the absorption spectrum is not clear
at present, but could be off-stoichiometry or disorder in the arrangement
of layered structure. We reiterate that the present infrared
absorption measurements have been carried out on a sample
characterised by sharp  x-ray diffraction pattern and
superconducting transition (cf. Fig.1).

The results of additional infrared absorption experiments in MgB$_2$, 
carried out over an extended range upto 2000 cm$^{-1}$, are shown
in Fig.3. The absorption spectrum is characterised by a linear background 
with broad humps centred at 485, 1040, 1442 and 1635 cm$^{-1}$. While the band 
at 1040 cm$^{-1}$ matches with crystalline B (also shown), we discount the 
possibilty of attributing this feature in MgB$_2$ to the presence of a second 
phase of B in our sample, since our x-ray diffraction pattern does not indicate
the presence of unreacted B. To substatntiate this, we also show in Fig. 3b  
the absorption spectrum of superconducting MgB$_2$ synthesised 
from amorphous B. This is also characterised by an absorption band 
at 1040 cm$^{-1}$, a feature that  is absent in the 
starting amorphous B. This could not have arisen due to crystallisation of
amorphous B itself, a process that is known  \cite{talley} to occur only beyond 
1500 K.  From these studies, as also several control infrared and x-ray diffraction 
experiments on MgB$_2$ samples in which intentionally crystalline 
B has been added, and studies \cite{bharathi} on Cu doped MgB$_2$, 
we infer that  the absorption spectrum of 
MgB$_2$ shown  Fig.3, viz., as characterised by a linear background with 
broad bands centred at 485, 1040, 1442 and 1635 cm$^{-1}$ is intrinsic to 
the system.  We note that the  absorption bands occurring at 1040, 
1442 and 1635 cm$^{-1}$ are
 beyond the range of optic phonons predicted by theoretical
calculations \cite{kortus,pickett,andersen,satta}. This once again points to the
important role of disorder in this system which may be activating the
the high frequency B-like modes \cite{nogi} in MgB$_2$. We also note that 
these high frequency features appear like replication of the absorption 
band at 485 cm$^{-1}$ and hence may be due to combination modes. The
relatively large intensity of these combination modes could be due to
the large anharmonicity, which has been shown to exist
\cite{yildrim} in this system.

\section{Temperature dependence of IR absorption}

The temperature dependence of IR absorption in the range of
300-650 cm$^{-1}$ has been followed across the superconducting
transition. Fig.4 summarises the results on the temperature
variation of the  frequencies and widths of the phonon features
at 333, 387 and 485 cm$^{-1}$. These  are seen to harden
initially with the lowering of temperature, but  this trend is
arrested at $\sim $ 100 K, below which they soften. An anomalous
variation  is also noticed in the widths, in that they increase
with the lowering of temperature. In contrast, it is seen from
Fig.5 that  the phonon modes at 542, 592 and 634 cm$^{-1}$,
which are B-like modes(cf. Fig.2), indicate a regular behaviour, viz., a
hardening of the modes and a decrease in phonon width with the
lowering of temperature.

The observed softening of the mode frequencies below $\sim $ 100
K in Fig.4 point to a structural anomaly. While the presence of an
incipient structural instability is well known in the case of
strong electron phonon coupled superconductors \cite{sinha},
this is not clearly established in the case of MgB$_{2}$. In
fact the experiments by Jorgensen et al \cite{jorgensen} on the
variation of lattice parameters of MgB$_{2}$ with temperature,
down to 11 K, indicate only a continuous decrease in the a and c
axis parameters with no anomalies. At the same time, experiments
on the suppression of superconductivity with small decrease in
c-parameter, obtained by Al doping \cite {slusky}, is indicative
of the fact that MgB$_2$ is near a structural instabilty. The
present infrared absorption measurements, which shows softening
of  the phonon modes suggest that there could be a minor
increase in c axis parameter below $\sim $100 K.  This calls for
further detailed structural investigations using techniques such
as EXAFS.

In the case of mode at 333 cm$^{-1}$, there is also a distinct
softening below  T$_{c}$. While the underlying reason for this
behaviour is not clear, we would like to point out that in the
experiments of Jorgensen et al \cite{jorgensen}, a distinct
increase in the Debye-Waller factor, U$_{33} (B)$, has been
observed below T$_{c}$.  This increase in the thermal factor of
B along the z axis may have a bearing on the softening of the
optic mode, observed in the present investigations. Earlier
neutron diffraction measurements by Sato et al \cite{sato}
indicated an anomolous behaviour of a mode at 17 meV,
subsequently discounted in other experiments \cite{yildrim}. In
the studies by Yildrim et al \cite{yildrim}, no substantial changes in the
phonon density of states has been observed with the lowering of
temperature from 200 to 7 K. While this is contrary to the
present observations, this may be  pointing  to the sensitivity
of infrared absorption measurements to pick up small changes in
the mode frequencies with the lowering of temperature. We also
note that while we have seen a small softening of the 333
cm$^{-1}$ mode, there is no corresponding change in the width,
which is expected to be modified with the appearence of
electronic energy gap in the superconducting phase
\cite{shirane}.

\section{conclusions}

To summarise, through infrared absorption measurements, optic
modes in MgB$_{2}$ have been identified. Infrared modes are seen
at 333 and 387 cm$^{-1}$, in accordance with theoretical
calculations \cite{kortus,pickett,andersen,satta}. While the
present experiments have been carried out on MgB$_2$ sample
characterised by sharp diffraction peaks and superconducting
transition, it is noted that the infrared absorption shows broad
features that can be accounted by invoking disorder and the
consequent relaxation of momentum selection rule.   The broad band
centred at 485 cm$^{-1}$, has been identified with the peak in
the phonon density of states and additional broad features
corresponding to combination modes are also seen in the present
experiments. In effect, the measured absorption spectrum is an
indication  of the phonon density of states as sampled by the
infrared absorption. These optic modes in the far infrared range
exhibit interesting temperature dependence involving a softening
below $\sim $ 100 K. The implications of this on a possible
structural anomaly needs to be investigated.

It is seen that  the infrared modes at 333 and 387 cm$^{-1}$ 
are characterised by a
widths of $\sim $ 40 cm$^{-1}$, which is considerably smaller
than the width of the Raman mode  \cite{bohnen,chen,goncharov}
$\sim $ 200 cm$^{-1}$. This may be taken as an evidence for the
strong electron phonon interaction with the E$_{2g}$ mode, as
has been suggested by theoretical calculations
\cite{pickett,andersen}.  However, given that in this system
disorder and dispersion effects seem to play a considerable
role, to dileneate the electronic contribution to the width and
to estimate the electron phonon interaction from the measured
widths, using Allen formula \cite{allen}, experiments need to be
carried out on single crystals of MgB$_2$. This will also help
to clarify the possible role, if any, of disorder on the
superconductivity in MgB$_{2}$.

\section{acknowledgements}
The authors thank Prof. A.K. Sood for discussions. 

\begin{figure}
\centerline{\psfig{figure=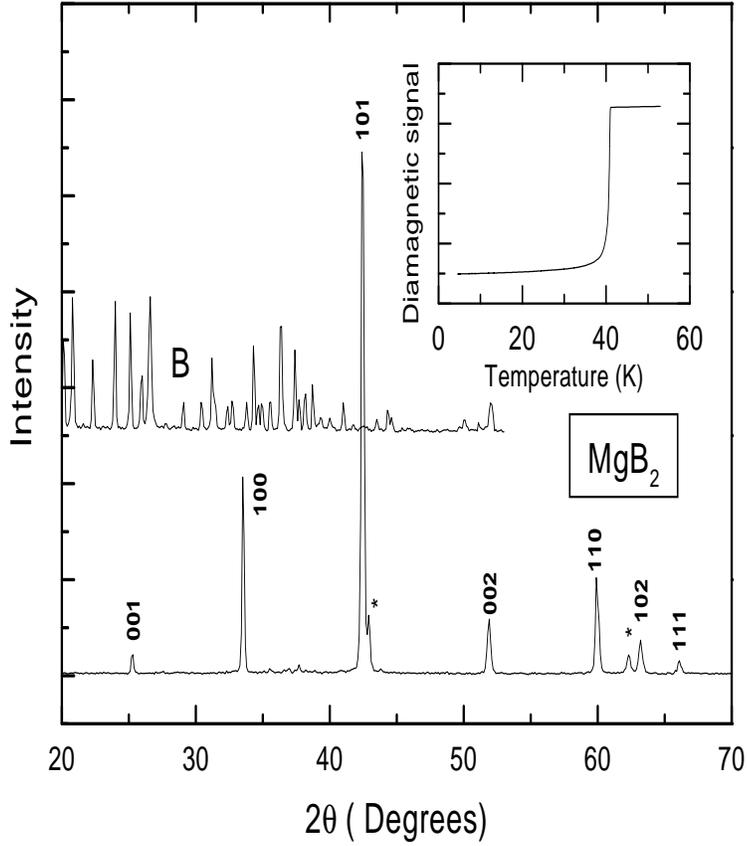,width=16cm,height=16cm}}
\caption{X-ray diffraction pattern of MgB$_2$ indexed to
hexagonal P6/mmm structure. The impurity lines due to MgO are
indicated by asterix. Also shown is the diffraction pattern of
crystalline B to indicate that no B features are discernable in the
diffraction pattern of MgB$_{2}$. The inset shows the
superconducting transition at 39 K.}
\label{Fig 1} 
\end{figure}

\begin{figure}
\centerline{\psfig{figure=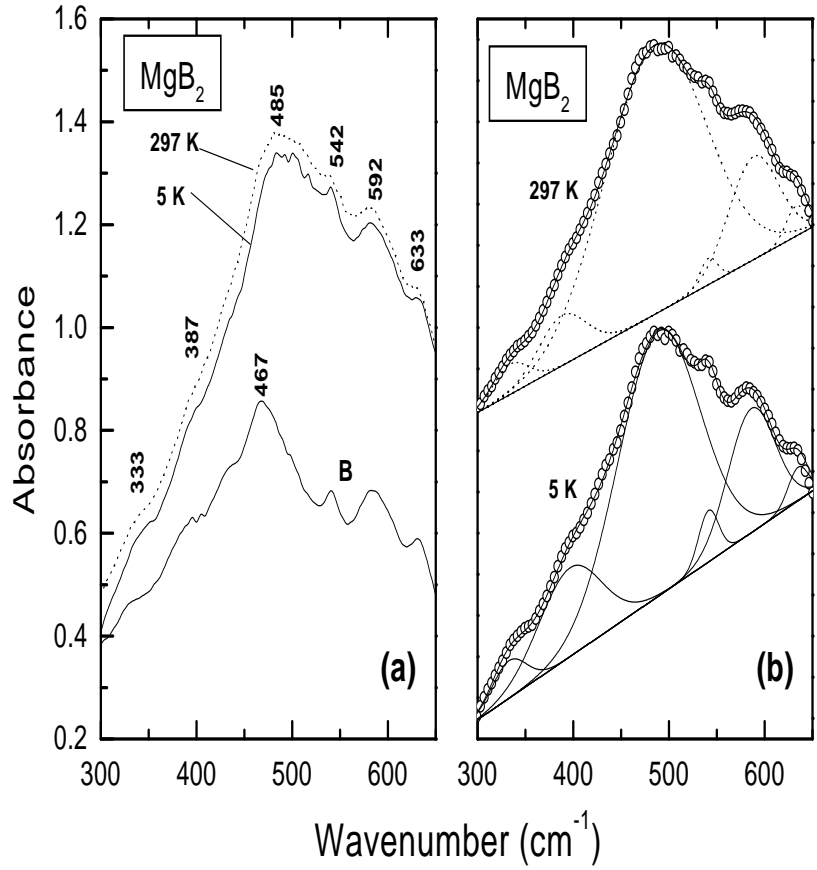,width=16cm,height=16cm}}
\caption{Panel (a) shows the absorption spectrum of MgB$_2$ at
297 and 5 K respectively. Also shown are the absorption spectra
of the crystalline B. Panel (b) shows the
fits of the absorption spectrum at 297 K and 5 K in terms of sum
of six Gaussians and a linear background. }
\label{Fig 2}
\end{figure}

\begin{figure}
\centerline{\psfig{figure=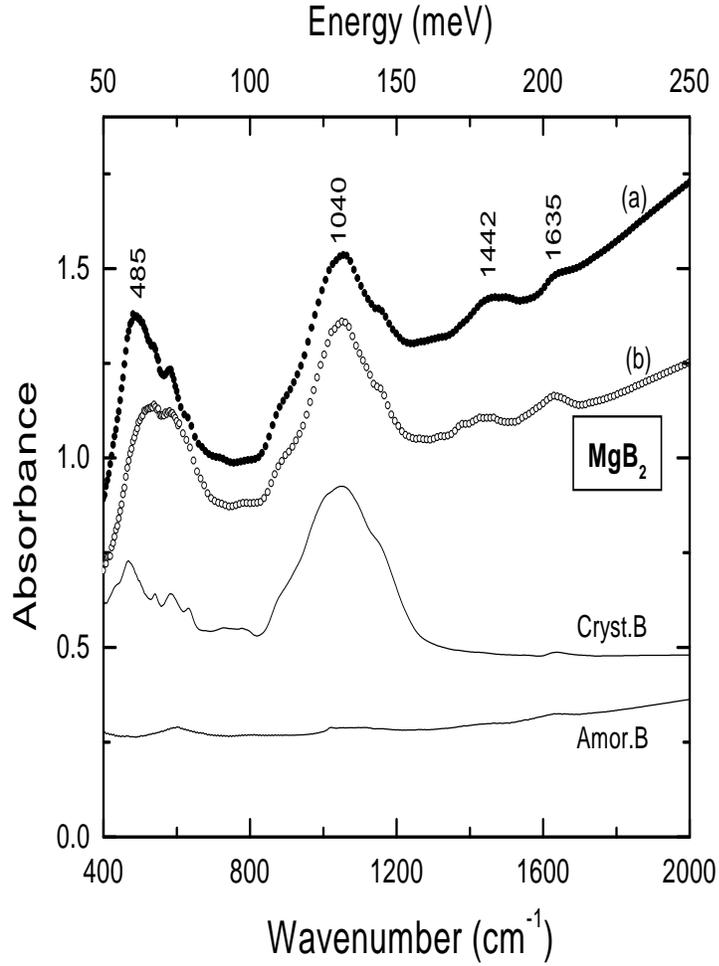,width=16cm,height=16cm}}
\caption{Mid-infrared absorption spectrum of superconducting MgB$_2$
(a) synthesised from crystalline  B, and (b) from amorphous B. For
comparison, the absorption spectra of the starting amorphous and
crystalline B are also shown. The absorption spectrum of MgB$_2$
is characterised by broad absorption bands at 485, 1040, 1442
and 1635 cm$^{-1}$.}
\label{Fig 3}
\end{figure}

\begin{figure}
\centerline{\psfig{figure=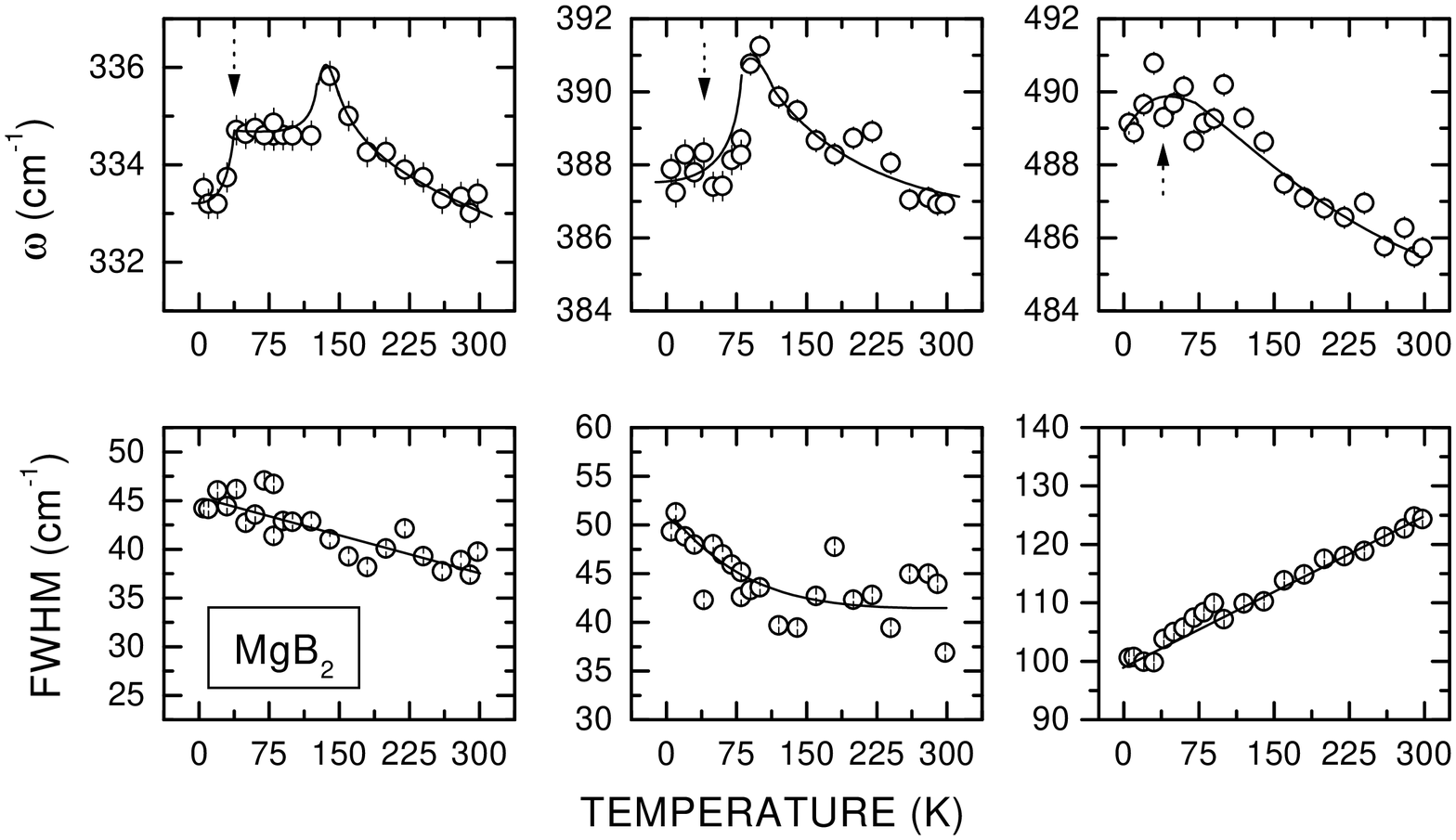,width=16cm,height=16cm}}
\caption{ Temperature dependence of the frequency and width of
the modes at 333, 387 and 485 cm$^{-1}$. The lines are guide to
the eye, and the arrow corresponds to the superconducting
transition.  Notice the softening of the infrared modes below
$\sim $100K  and a distinctive softening below T$_c$ for the 333
cm$^{-1}$ mode. }
\label{Fig 4}
\end{figure}

\begin{figure}
\centerline{\psfig{figure=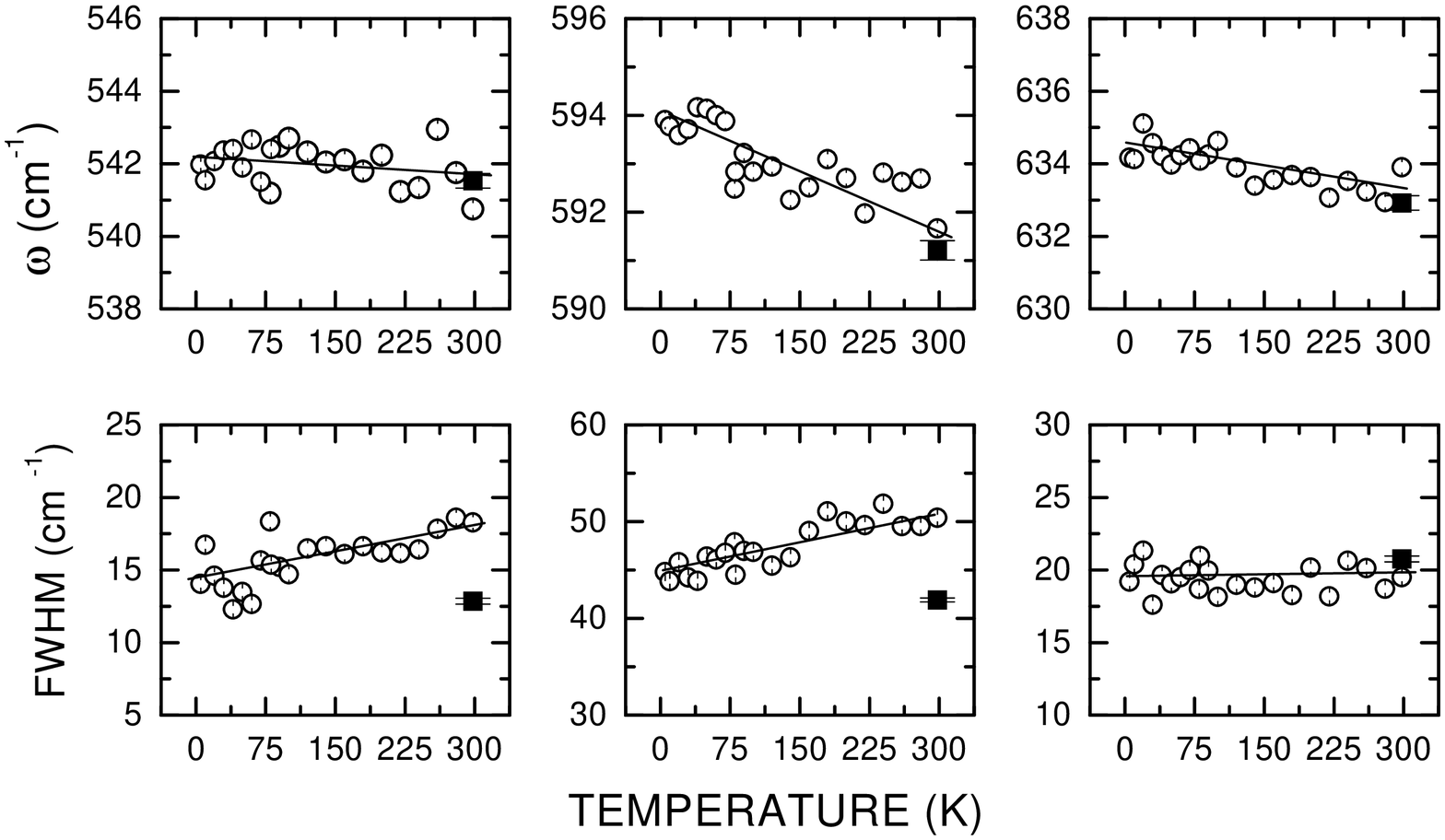,width=16cm,height=16cm}}
\caption{Temperature dependence of the frequency and
width of the modes at 542, 594 and 634 cm$^{-1}$. In
contrast to the modes shown in Fig.4, these show a regular
hardening behaviour with the lowering of temperature. The filled
squares are the mode frequency and width for crystalline B.
The lines are guide to the eye.  }
\label{Fig 5}
\end{figure}

\end{document}